\documentclass[aps, prd, twocolumn, nofootinbib]{revtex4-2}

\usepackage[colorlinks=true, linkcolor=blue, citecolor=blue, urlcolor=blue]{hyperref}
\usepackage{amsmath, graphicx, braket, orcidlink}

\usepackage[dvipsnames]{xcolor}

\begin{document}

\title{Sphalerogenesis}

\author{Masanori Tanaka~\orcidlink{0000-0002-1303-7043}}
\email{tanaka@pku.edu.cn}
\affiliation{Center for High Energy Physics, Peking University, Beijing 100871, China}

\begin{abstract}

We propose a mechanism called \textit{sphalerogenesis} to explain the baryon asymmetry of the universe (BAU).
The BAU is explained by a CP‐violating decay of the electroweak sphaleron. 
We introduce a dimension-six operator constructed from weak gauge fields: $Q_{\widetilde{W}} \sim \Lambda^{-2} \epsilon_{ijk} \widetilde{W}_{\mu \nu}^{i} W^{j \nu \rho} W_{\rho}^{k\mu}$. 
We find that the BAU can be explained if $\Lambda \simeq 38\,{\rm TeV}$.
This scenario can be tested by electron electric dipole moment measurements in the near future.

\end{abstract}

\maketitle

\section{Introduction}

One of the fundamental open questions in particle cosmology is the origin of the baryon asymmetry of the universe (BAU)~\cite{Planck:2018vyg}. 
The observed baryon-to-entropy ratio is constrained as~\cite{Planck:2018vyg, ParticleDataGroup:2024cfk}
\begin{align}
\label{eq:BAU}
8.41 \times 10^{-11} < \frac{n_{B}}{s} < 8.75 \times 10^{-11} \,.
\end{align}
Even if our universe has a large initial baryon asymmetry before cosmological inflation, the observed BAU cannot be consistently explained~\cite{Murai:2023ntj}.
Therefore, a dynamical mechanism that satisfies Sakharov's conditions is required to produce the BAU after inflation~\cite{Sakharov:1967dj}.

Electroweak baryogenesis in the Standard Model (SM) was a plausible candidate~\cite{Kuzmin:1985mm}.
However, lattice simulations have firmly established that the electroweak (EW) phase transition in the SM is a smooth crossover rather than being first order~\cite{Kajantie:1996mn,Csikor:1998eu,Aoki:1999fi,DOnofrio:2014rug,DOnofrio:2015gop}, which is required to satisfy the third Sakharov's condition~\cite{Kuzmin:1985mm}. 
Moreover, it has been confirmed that the CP violation in the Cabibbo-Kobayashi-Maskawa (CKM) matrix is far too small to generate the observed asymmetry~\cite{Gavela:1994dt, Huet:1994jb}. 
These negative results motivate physics beyond the SM.

EW sphalerons play an important role in various baryogenesis scenarios because their energy is related to the transition rate of the baryon number violating process~\cite{Kuzmin:1985mm}. 
The EW sphaleron, which corresponds to the saddle point of the energy functional between topologically distinct vacua, is a non-perturbative solution to the classical field equations for the $SU(2)_{L}$ gauge and Higgs doublet fields~\cite{Manton:1983nd}. 
In Ref.~\cite{Manton:1983nd}, the EW sphaleron solution in the SM has been obtained by considering a non-contractive loop which connects adjacent vacua. 
In the early universe, the thermal transition between topologically distinct vacua can occur via the EW sphaleron.
This process is called the EW sphaleron process, which changes the number of baryons and leptons~\cite{tHooft:1976rip}.

Recently, a new possibility has been suggested to explain the BAU without extending the SM by using the CP asymmetry in the EW sphaleron process~\cite{Kharzeev:2019rsy}. 
In the new scenario~\cite{Kharzeev:2019rsy}, all Sakharov's conditions can be simultaneously satisfied via the CP-violating fermion emission from the sphaleron explosion. 
The EW sphaleron process occurs dominantly along the path following the non-contractive loop discussed in Ref.~\cite{Manton:1983nd}.  
In the early universe, thermal fluctuations cause deviations from this path, resulting in the sphaleron-like transition~\cite{Ostrovsky:2002cg, Flambaum:2010fp, Kharzeev:2019rsy, Hong:2023zrf}.
In general, information about such sphaleron-like transitions is included in the prefactor of the full sphaleron transition rate as a fluctuation effect~\cite{Arnold:1987mh, Hong:2023zrf}.  
Such sphaleron-like transitions are energetically less favorable than the true sphaleron transition, so they gradually decouple earlier than the latter as the universe cools down. 
Then, the third Sakharov's condition can be satisfied during this decoupling process instead of the first-order EW phase transition. 
In Ref.~\cite{Kharzeev:2019rsy}, it has been discussed that incorporating this effect may amplify the CP asymmetry from the CKM phase, allowing for the explanation of sufficient baryon numbers.  
The magnitude of CP asymmetry in the EW sphaleron process has been estimated by comparing emission rates for particles and anti-particles from the EW sphaleron process in Ref.~\cite{Kharzeev:2019rsy}. 
Using the result, they concluded that the BAU may be explained without extending the SM. 
However, it has been pointed out that the evaluation of the baryon asymmetry in Ref.~\cite{Kharzeev:2019rsy} does not include the washout effect of the EW sphaleron itself~\cite{Hong:2023zrf}. 
Taking into account the effect, the prediction of $n_{B}/s$ is reduced to $10^{-14}$, well below the observed value in Eq.\,\eqref{eq:BAU}.
Thus, new physics with CP-violating sources is still required to explain the BAU. 

Although the scenario in Ref.~\cite{Kharzeev:2019rsy} fails to reproduce the observed BAU within the SM, its underlying mechanism remains viable in extensions beyond the SM. 
However, estimating the CP asymmetry in the EW sphaleron process is technically challenging in general new physics frameworks. 
An alternative method for estimating this CP asymmetry was proposed in Refs.~\cite{Nauta:2000xi, Nauta:2002ru}.
These works have shown that the CP asymmetry arises dynamically when CP-violating dimension-eight operators are introduced. 
This dynamical CP asymmetry is obtained by treating the Chern–Simons number as a dynamical variable in the sphaleron transition~\cite{Funakubo:1991hm, Funakubo:1992nq, Tye:2015tva, Tye:2016pxi, Tye:2017hfv, Funakubo:2016xgd, Qiu:2018wfb}. 
However, the author of Ref.~\cite{Nauta:2002ru} concluded that explaining the BAU is difficult because the decoupling of sphaleron-like processes has not been considered in their calculations. 

In this paper, we build on the mechanism in Ref.~\cite{Kharzeev:2019rsy} and combine it with the analyses of Refs.~\cite{Nauta:2000xi, Nauta:2002ru} to reproduce the observed BAU in Eq.\,\eqref{eq:BAU}. 
We refer to this baryogenesis mechanism as \textit{sphalerogenesis}.
To obtain model‐independent predictions, we work within the Standard Model Effective Field Theory (SMEFT) in the Warsaw basis~\cite{Grzadkowski:2010es}, extending the SM Lagrangian by a dimension‐six operator shown in Eq.\,\eqref{eq:Scp}. 
This operator, not previously considered in Refs.~\cite{Nauta:2000xi, Nauta:2002ru}, can in fact dominate the CP violation in the EW sphaleron process. 
We demonstrate that this scenario can reproduce the observed BAU while satisfying the current strongest limits from electron electric dipole moment (EDM) measurements by JILA~\cite{Roussy:2022cmp}.

\section{CP asymmetry in EW sphaleron process}

Here we discuss the CP asymmetry in the EW sphaleron process.
The action we consider is defined by 
\begin{align}
\label{eq:tot_action}
S = S_{\rm SM} + S_{\rm CP} \,,
\end{align}
where $S_{\rm SM}$ is the SM action, which is given by
\begin{align}
\label{eq:Ssm}
S_{\rm SM} = \int d^4x \left[ - \frac{1}{2} {\rm tr} \left( W_{\mu \nu} W^{\mu \nu} \right) + |D_{\mu} \Phi|^2 - V(\Phi) \right] \,, 
\end{align}
where $W_{\mu \nu}$ and $\Phi$ are the $SU(2)_{L}$ gauge field strength and the SM Higgs doublet field, respectively. 
The Higgs potential $V(\Phi)$ is given by
\begin{align}
V(\Phi) = - \mu^2 |\Phi|^2 + \lambda |\Phi|^4 \,. 
\end{align}
Since the contribution from $U(1)_{Y}$ gauge fields to the sphaleron properties is only a few percent~\cite{Klinkhamer:1984di, Klinkhamer:1990fi}, we do not take it into account.
The term $S_{\rm CP}$ is defined as~\cite{Grzadkowski:2010es}
\begin{align}
\label{eq:Scp}
S_{\rm CP} = - \int d^4x \frac{g}{3 \Lambda^2} \epsilon_{ijk} \widetilde{W}^{i}_{\mu \nu} W^{j\nu \rho} W^{k\mu}_{\rho} \,, 
\end{align}
with $\widetilde{W}^{j \mu \nu} = \frac{1}{2} \epsilon^{\mu \nu \rho \sigma} W_{\rho \sigma}^{j}$. 
The Latin and Greek indices represent the $SU(2)$ index and spacetime coordinates, respectively. 
The factor $g$ denotes the $SU(2)$ gauge coupling. 
This term can be a source of CP asymmetry in the EW sphaleron process. 
For the value of $g^2$ and $\lambda$ at finite temperatures, we use $\left. g^2 \right|_{T\neq0} \simeq 0.39$ and $\left. \lambda/g^2 \right|_{T\neq0} \simeq 0.22$ obtained by extrapolating from the three-dimensional effective field theory as taken in Ref.~\cite{Hong:2023zrf}. 

To estimate the sphaleron action, we adopt the following sphaleron ansatz~\cite{Manton:1983nd}
\begin{align}
\label{eq:sph_ansatz}
\begin{aligned}
W_{\mu}(\mu, r, \theta, \phi) dx^{\mu} 
=& - \frac{i}{g} f(r) d U^{\infty} U^{-1} \,,  \\
\Phi(\mu , r, \theta, \phi) 
=& \frac{v}{\sqrt{2}} \left[ 1- h(r) \right] \left( \begin{array}{c} 0 \\ e^{-i\mu} c_{\mu} \end{array} \right) \\ 
& +
\frac{v}{\sqrt{2}} h (r) U^{\infty} \ \left( \begin{array}{c} 0 \\ 1 \end{array} \right) \,, 
\end{aligned}
\end{align}
with
\begin{align}
  U^{\infty} =
  \left( \begin{array}{cc} 
  e^{i \mu} (c_{\mu} - i s_{\mu} c_{\theta}) & e^{i \phi} s_{\mu} s_{\theta} \\
  - e^{-i \phi} s_{\mu} s_{\theta} & e^{-i\mu} (c_{\mu} + i s_{\mu} c_{\theta})
  \end{array} \right) \,,
\end{align}
where $s_{x} \equiv \sin x$ and $c_{x} \equiv \cos x$. 
The parameters $\mu \in [0, \pi]$ and $(r, \theta, \phi)$ characterize the non-contractive loop and the spherical coordinates, respectively. 
The factor $v$ represents the vacuum expectation value of the SM Higgs field $\braket{\Phi} = \frac{1}{\sqrt{2}}(0, v)^{\rm T}$. 
For the temperature dependence of $v$, we employ the fitting function obtained by the lattice simulation in the SM: $v(T) \simeq 3T \sqrt{1-T/T_{\rm EW}}$, where $T_{\rm EW} = 159.5 \pm 1.5 \,{\rm GeV}$~\cite{DOnofrio:2014rug, Hong:2023zrf}. 
The profile functions $f(r)$ and $h(r)$ should satisfy the following boundary condition
\begin{align}
&\lim_{r \to \infty} f(r) = \lim_{r \to \infty} h(r) = 1 \,, \\
&\lim_{r \to 0} f(r) = \lim_{r \to 0} h(r) = 0 \,. 
\end{align}
The precise form of $f(r)$ and $h(r)$ in the SM has been discussed in the literature (see, e.g., Ref.~\cite{Kanemura:2020yyr}). 
Instead of the numerical solution, we employ the ansatz for the profile functions $f^{b}(r)$ and $h^{b}(r)$ proposed in Ref.~\cite{Klinkhamer:1984di}. 
The ansatz is parameterized by parameters $\Xi$ and $\Omega$ that characterize the size of the EW sphaleron. 
We find that this configuration becomes a saddle point when $\Xi = 1.467$ and $\Omega = 1.701$, which is consistent with Ref.~\cite{Hong:2023zrf}.
In the following, we call the field configuration with $\Xi = 1.467 \, (= \Xi_{0})$ and $\Omega = 1.701 \,( = \Omega_{0})$ as the true sphaleron. 

By using the ansatz for the EW sphaleron in Eq.\,\eqref{eq:sph_ansatz}, the action in Eq.\,\eqref{eq:tot_action} is given by 
\begin{align}
\label{eq:action_sph}
S = \int d \eta \left[ \frac{M(\mu)}{2} (\mu')^2 + G(\mu) (\mu')^3 - V(\mu) \right] \,, 
\end{align}
where $\eta  = g v t$ and $\mu' \equiv  d\mu/d\eta$.
The explicit form of Eq.\,\eqref{eq:action_sph} is shown in Appendix~\ref{app:sphaleron_action}.
In deriving the action in Eq.\,\eqref{eq:action_sph}, we have dropped total time-derivative terms. 
The functions $M(\mu), G(\mu)$ and $V(\mu)$ are given by 
\begin{align}
\label{eq:Mmu}
&M(\mu) = \frac{4\pi}{g^2} \left( \alpha_{0} + \alpha_{1} c_{\mu}^2 + \alpha_{2} c_{\mu}^4  \right) \,, \\
&G(\mu) = \frac{256 \pi}{45} s_{\mu}^2 (4 - s_{\mu}^2) \left( \frac{v}{\Lambda} \right)^2 \,, \\
&V(\mu) = \frac{4\pi}{g^2} s_{\mu}^2 \left( \beta_{1} + \beta_{2} s_{\mu}^2 \right) \,. 
\end{align}
These agree with Ref.~\cite{Funakubo:2016xgd} except for $G(\mu)$. 
The coefficients $\alpha_{i} \, (i = 0,1,2) $ and $ \beta_{i} \, (i=1,2)$ are determined numerically from the profile functions $f(r)$ and $h(r)$. 
We have numerically checked that the values of $\alpha_{i}$ and $\beta_{i}$ we obtained are consistent with the results in Ref.~\cite{Funakubo:2016xgd}.
Interestingly, the function $G(\mu)$ does not depend on the form of the profile functions.

One might expect that other CP-violating dimension-six operators such as $|\Phi|^2 {\rm tr}(W_{\mu \nu} \widetilde{W}^{\mu \nu})$ may be sources of CP asymmetry in the EW sphaleron process. 
As a contrast, in the tachyonic EW phase transition, the evolution of the Higgs field results in the charge potential for the Chern-Simons number~\cite{Shaposhnikov:1987pf, Garcia-Bellido:1999xos, Turok:1990in, Tranberg:2003gi, Garcia-Bellido:2003wva, Tranberg:2009de}.
On the other hand, in our setup, such operators can be expressed as a total time derivative because the time evolution of the Higgs and $SU(2)$ gauge fields is specified by the ansatz in Eq.\,\eqref{eq:sph_ansatz}~\cite{Nauta:2002ru, Gan:2017mcv}.
As a result, CP-violating bosonic dimension-six operators such as $|\Phi|^2 {\rm tr}(W_{\mu \nu} \widetilde{W}^{\mu \nu})$ do not induce CP violation in the EW sphaleron process.
For this reason, the dimension-eight operators have been considered to produce the CP asymmetry in the EW sphaleron process in Refs.~\cite{Nauta:2000xi, Nauta:2002ru}. 
However, we confirm that the operator in Eq.\,\eqref{eq:Scp} can have a non-trivial effect on the sphaleron process even if its mass dimension is six. 
Therefore, we find that the operator in Eq.\,\eqref{eq:Scp} can provide a dominant CP-violating contribution in our setup.

Here we use a formalism where the loop parameter $\mu$ is regarded as a dynamical variable~\cite{Funakubo:1991hm, Funakubo:1992nq, Tye:2015tva, Tye:2016pxi, Tye:2017hfv, Funakubo:2016xgd, Qiu:2018wfb}. 
Because of the spherical symmetry in the sphaleron ansatz in Eq.~\eqref{eq:sph_ansatz}, we can regard Eq.~\eqref{eq:action_sph} after integrating over spatial coordinates as the one-dimensional (in time) effective action for the sphaleron process, known as the reduced model~\cite{Funakubo:1992nq}.
Defining the dynamical variable as $Q = \mu/(gv)$, the corresponding Lagrangian reads~\cite{Nauta:2002ru}
\begin{align}
L = \frac{M(Q)}{2} \left( \frac{d Q}{d t} \right)^2 + G(Q) \left( \frac{d Q}{dt} \right)^3 - V(Q) \,.
\end{align}

It should be emphasized that extrapolating the effective Lagrangian to arbitrarily large values of $\dot{Q}$ lies outside its regime of validity, since the term proportional to $\dot{Q}^3$ would eventually dominate in that region.
On the other hand, as shown in Ref.~\cite{Nauta:2002ru}, the typical value of $\dot{Q}$ relevant for a sphaleron transition is estimated as $\dot{Q} \sim \sqrt{T/M_{\rm sph}}\,,$ where \(M_{\rm sph}\) is the sphaleron mass, defined by $M_{\rm sph} \equiv gv\,M(\mu=\pi/2)\sim 92~{\rm TeV}$.
Since sphalerogenesis takes place at temperatures of the order of $T\sim 100~{\rm GeV}$, the relevant velocities remain sufficiently small.
In our analysis, we further assume that individual sphaleron processes occur independently.
In principle, if multi-sphaleron processes involving large simultaneous changes in $Q$ are important, $\dot{Q}$ could be enhanced through a chain-like sequence of multiple sphaleron transitions.
However, such multi-sphaleron processes are expected to be suppressed because multi-sphaleron configurations have substantially larger energies and hence much smaller transition rates~\cite{Kleihaus:1994yj}.
We therefore expect the independent-process approximation to be reasonable. 
Under this assumption, higher-order terms in $\dot{Q}$ remain subleading and can be treated as perturbative corrections.

The canonical conjugate momentum of $Q$ is defined by 
\begin{align}
\pi_{Q} = \frac{\partial L}{\partial \dot{Q}} = M(Q) \dot{Q} + 3 G(Q) \dot{Q}^2 \,. 
\end{align}
Using $\pi_{Q}$, the Hamiltonian is obtained as
\begin{align}
\label{eq:hamiltonian}
\mathcal{H} &= \pi_{Q} \dot{Q} -L \nonumber \\
&= \frac{1}{2 M(Q)} \pi_{Q}^2 - \frac{G(Q)}{M^3(Q)} \pi_{Q}^3 + V(Q) \,,
\end{align}
where we neglect $O(G^2)$ terms, which are of the same-order contributions as dimension-eight operators. 
Since Eq.\,\eqref{eq:hamiltonian} contains a cubic term in $\pi_{Q}$, the transition rates along the positive and negative Chern–Simons directions differ.
To estimate the difference, we calculate the mean velocity along the positive and negative Chern-Simons number directions at fixed $\pm \overline{Q}$.
The mean velocity at $\pm \overline{Q}$ is given by~\cite{Nauta:2000xi, Nauta:2002ru}
\begin{align}
v_{\pm}(\overline{Q}) = \frac{\braket{\dot{Q} \Theta(\pm \dot{Q}) \delta(Q \mp \overline{Q})}}{\braket{\Theta(\pm\dot{Q}) \delta(Q \mp \overline{Q})}} \,, 
\end{align}
where $\Theta(x)$ and $\delta(x)$ are the Heaviside function and delta function, respectively. 
In our analysis, we take $\overline{Q} = \pi/(2 gv)$, which corresponds to the sphaleron configuration. 
The braket denotes a thermal average, which is defined for a physical quantity $\mathcal{O}$ as
\begin{align}
\braket{\mathcal{O}} \equiv \frac{1}{Z_{0}} \iint d \pi_{Q} dQ \mathcal{O} e^{- \mathcal{H}/T} \,, 
\end{align} 
where $Z_{0}$ is the normalization factor. 
Due to the cubic term in Eq.\,\eqref{eq:hamiltonian}, $v_{+}(\overline{Q})$ and $v_{-}(\overline{Q})$ are not the same. 
Therefore, we define a measure of the CP asymmetry in the EW sphaleron process as
\begin{align}
\label{eq:Acp}
A_{\rm CP} = \frac{v_{+}(\overline{Q}) - v_{-}(\overline{Q})}{v_{+}(\overline{Q}) + v_{-}(\overline{Q})} \,. 
\end{align}

In Ref.~\cite{Kharzeev:2019rsy}, the CP violation in the EW sphaleron process has been estimated by considering the interaction between the gauge field configuration of the sphaleron and fermions produced via sphaleron explosions. 
On the other hand, our analysis does not rely on the sphaleron explosions, but on the asymmetry of sphaleron transition probabilities over the sphaleron potential in the thermally populated ensemble.
We expect that this method is suitable for estimating the CP asymmetry during the gradual decoupling of the EW sphaleron process.

\section{Boltzmann equation for the baryon number}

We introduce the Boltzmann equation for the baryon number in the scenario of sphalerogenesis. 
In our scenario, the baryon asymmetry is produced during the gradual sphaleron decoupling. 
To describe the decoupling process, we employ the formalism developed in Ref.~\cite{Hong:2023zrf}. 
In their formalism, the sphaleron transition rate obtained by the lattice simulation is decomposed into contributions from sphaleron-like configurations as~\cite{Hong:2023zrf}
\begin{align}
\Gamma_{\rm sph}^{\rm lattice}(T) = \int da \Gamma_{\rm sph}(a, T) \,, 
\end{align}
with 
\begin{align}
\label{eq:Gamma_sph}
\Gamma_{\rm sph}(a, T)= \frac{e^{-E_{\rm sph}(a, T)/T}}{\int da' e^{- E_{\rm sph}(a', T)/T}} \Gamma_{\rm sph}^{\rm lattice}(T) \,,
\end{align}
where $\Gamma_{\rm sph}^{\rm lattice}(T)$ is the sphaleron transition rate obtained by lattice simulations~\cite{DOnofrio:2014rug}. 
The parameter $a$ characterizes the size of the sphaleron-like field configuration. 
The sphaleron-like field configuration is given by the profile functions $f^{b}(r)$ and $h^{b}(r)$ with $\Xi = a \Xi_{0}$ and $\Omega = a \Omega_{0}$~\cite{Hong:2023zrf}. 
Thus, the true sphaleron corresponds to the case with $a= 1$. 
For the sphaleron-like configuration, $M(\mu)$ and $V(\mu)$ depend on $a$:~$M(a, \mu)$ and $V(a, \mu)$. 
In this case, the energy of the sphaleron-like configuration is given by $E_{\rm sph}(a, T) = g v(T) V(a, \mu=\pi/2)$. 

Evaluating Eq.\,\eqref{eq:hamiltonian} at $\overline{Q} = \pi/(2 gv)$ and using $M(\mu ,a)$, we obtain
\begin{align}
\label{eq:Acp_sph}
A_{\rm CP}(a, T) = \sqrt{ \frac{8T}{gv(T)} } \frac{ G(\mu=\pi/2, T)}{[ M(a, \mu=\pi/2)]^{3/2} } \,. 
\end{align}
We emphasize that Eq.\,\eqref{eq:Acp_sph} encodes the dependence on $a$ and $T$ in contrast to Ref.~\cite{Kharzeev:2019rsy}. 
If $A_{\rm CP}(a, T)$ takes a non-zero value, there is an asymmetry between transitions mediated by sphaleron-like and antisphaleron-like configurations with the same size $a$ and corresponding energy $E_{\rm sph}(a, T)$. 
We therefore use $A_{\rm CP}(a, T)$ as an approximate measure of the CP-odd bias in the EW sphaleron transition rates.

Since $E_{\rm sph}(a, \mu) \geq E_{\rm sph}(a =1, \mu)$, the decoupling of sphaleron-like configurations with $a \neq 1$ occurs earlier than that of the true sphaleron.
The decoupling temperature for the sphaleron with the size $a$ is determined by 
\begin{align}
\label{eq:Gsph_H}
\frac{\Gamma_{\rm sph}(a, T_{*}(a))}{T_{*}^3(a)} = c H(T_{*}(a)) \,, 
\end{align}
where $H$ denotes the Hubble parameter. 
The factor $c$ parametrizes the ambiguity of the decoupling process.

\begin{figure}[t]
\centering
\includegraphics[width=0.98\linewidth]{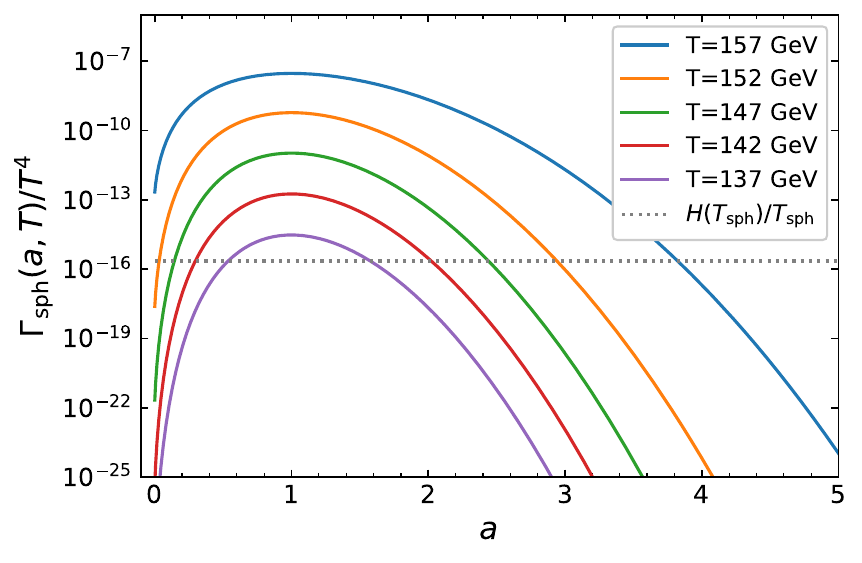}
\caption{
The temperature dependence of the sphaleron-like transition rate.
The gray dotted line is the ratio $H(T_{\rm sph})/T_{\rm sph}$, where $T_{\rm sph}$ is the sphaleron decoupling temperature at which the EW sphaleron process is completely decoupled.
}
\label{fig:Gsph_T}
\end{figure}

In Figure~\ref{fig:Gsph_T}, the temperature dependence of the sphaleron-like transition rate is shown. 
As the temperature decreases, the sphaleron-like configurations that are too small and too large gradually decouple.
Since this sphaleron decoupling process realizes a non-equilibrium environment, the third Sakharov's condition is satisfied during this process.
Based on Figure~\ref{fig:Gsph_T}, we can determine the threshold size of such sphaleron-like configurations using the condition in Eq.\,\eqref{eq:Gsph_H} for each temperature.
In our analysis, we express the threshold value for small and large sphalerons as $a_{l}(T)$ and $a_{u}(T)$ at temperature $T$, respectively. 

According to Fig.~\ref{fig:Gsph_T}, sphaleron-like configurations give two independent contributions to baryon asymmetry production, depending on their size $a$. 
In the case with $a_{l}(T) < a < a_{u}(T)$, such sphaleron-like configurations washout the produced baryon asymmetry because those configurations can be still actively produced due to the thermal fluctuation. 
On the other hand, for sphaleron-like configurations with $a_{l}(T) > a$ or $a > a_{u}(T)$, those immediately decay and their production from the thermal bath is decoupled, as shown in Fig.~\ref{fig:Gsph_T}. 
Therefore, if there is a CP asymmetry in the decay process of the sphaleron-like field configuration, the baryon asymmetry can be produced via the decoupling (or decay) of large and small sphaleron-like configurations. 
The corresponding Boltzmann equation for the baryon number density $n_{B}$ is expressed as~\cite{Joyce:1994fu, Bochkarev:1987wf, Joyce:1994zn, Hong:2023zrf}
\begin{align}
\label{eq:boltzmann_eq_t}
\frac{d n_{B}}{dt} + 3 H(t) n_{B} = - \Gamma_{B}(t) n_{B} + P(t) \,, 
\end{align}
where $\Gamma_{B}(t)$ and $P(t)$ represent the washout and source terms, respectively. 
The explicit expression of $\Gamma_{B}(t)$ and $P(t)$ is shown in Appendix~\ref{sec:BoltzmannEq}. 
In this analysis, we assume that the sphaleron-like configurations within the region $a_{l}(T) < a < a_{u}(T)$ ($a_{l}(T) > a$ or $a > a_{u}(T)$) contribute to $\Gamma_{B}(t)~(P(t))$. 
The second term in Eq.~\eqref{eq:boltzmann_eq_t} denotes the dilution effect due to the cosmological expansion. 
Although this contribution is often ignored in EW baryogenesis~\cite{Joyce:1994fu, Bochkarev:1987wf, Joyce:1994zn}, it cannot be neglected in sphalerogenesis because the sphaleron decoupling occurs over a long period as shown in Fig.~\ref{fig:Gsph_T}. 
In EW baryogenesis, the source term in Eq.~\eqref{eq:boltzmann_eq_t} depends on the charge potential for the baryon number around vacuum bubble walls~\cite{Joyce:1994fu, Bochkarev:1987wf, Joyce:1994zn}. 
In sphalerogenesis, the source term is produced via the CP violation in the decay process of the sphaleron-like field configuration. 
Employing the adiabatic time-temperature relation $dT/dt = - HT$, the Boltzmann equation is expressed in terms of the temperature as~\cite{Hong:2023zrf}
\begin{align}
\label{eq:boltzmann_eq}
- H T \frac{d n_{B}}{dT} + 3 H n_{B} = - \Gamma_{B}(T) n_{B} + P(T) \,. 
\end{align}
Solving this equation with $A_{\rm CP}$ given in Eq.~\eqref{eq:Acp_sph}, we can obtain a prediction for the baryon asymmetry. 
The numerical solution of Eq.~\eqref{eq:boltzmann_eq} is shown in Appendix~\ref{sec:BoltzmannEq}.

\section{Numerical results}

Here we present our numerical results for the predicted baryon asymmetry based on the discussion above.

\begin{figure}
\centering
\includegraphics[width=0.98\linewidth]{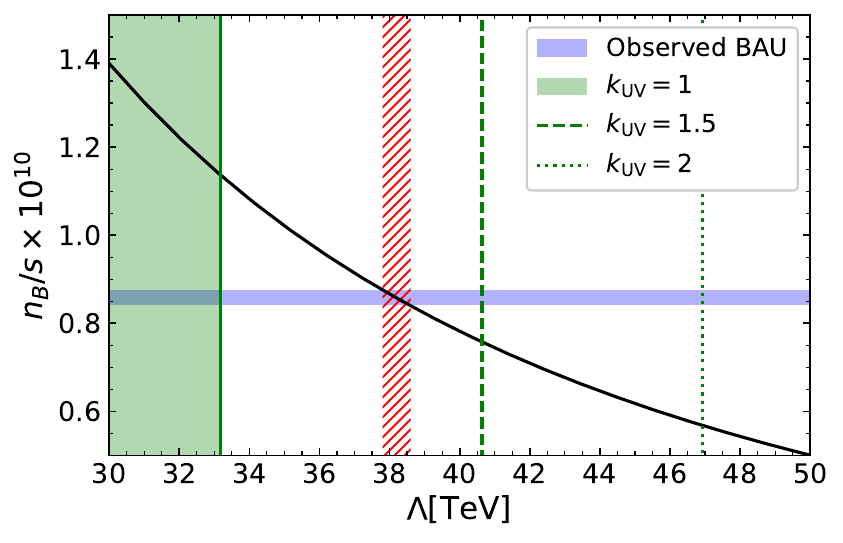}
\caption{
Prediction on the BAU for each $\Lambda$ with $c = 1$.
The black solid line indicates the prediction on the baryon asymmetry for each $\Lambda$. 
For the value of $\Lambda$ in the region hatched in red, the observed BAU can be explained. 
The green lines show the current EDM constraint by JILA~\cite{Roussy:2022cmp} for each $k_{\rm UV}$ value.
}
\label{fig:summary}
\end{figure}

Figure~\ref{fig:summary} shows the cutoff scale $\Lambda$ in Eq.\,\eqref{eq:Scp} required to explain the BAU. 
Fig.\,\ref{fig:summary} indicates that the observed baryon asymmetry in Eq.\,\eqref{eq:BAU} is reproduced if $\Lambda \simeq 38\,{\rm TeV}$.
This parameter region can evade current constraints from electron EDM measurements~\cite{Roussy:2022cmp}. 
In the presence of the operator in Eq.\,\eqref{eq:Scp}, it is expected that the electron EDM deviates from the SM prediction~\cite{Panico:2018hal,Dekens:2019ept,Kley:2021yhn,Abe:2024mwa,Banno:2024apv}. 
Neglecting the SM contribution, the predicted electron EDM $d_{e}$ normalized by the electric charge $e$ is given by~\cite{Banno:2024apv}
\begin{align}
\label{eq:electron_de}
\frac{d_{e}}{e} =\frac{k_{\rm UV} g^2 m_{e}}{96\pi^2 \Lambda^2} = 4.1 \times 10^{-30} {\rm cm} \,  \left( \frac{k_{\rm UV}}{1.35}\right) \left( \frac{38\,{\rm TeV}}{\Lambda} \right)^2 \,, 
\end{align}
where the factor $k_{\rm UV}$ parametrizes an $O(1)$ ambiguity due to evanescent operators~\cite{Kley:2021yhn, Banno:2024apv}.
On the other hand, the current strongest constraint from the electron EDM measurement in JILA is given by~\cite{Roussy:2022cmp}
\begin{align}
\label{eq:JILA}
\left| \frac{d_{e}}{e} \right| < 4.1 \times 10^{-30} {\rm cm}\,.
\end{align}
In Fig.\,\ref{fig:summary}, the current constraint on $\Lambda$ from the JILA results is shown as the green lines for each $k_{\rm UV}$ value.
In the case with $k_{\rm UV} = 1$ (green solid line), our scenario can be tested if future electron EDM measurements reach a sensitivity at the level of $|d_{e}/e|  < 3 \times 10^{-30}\,{\rm cm}$. 
Since higher accuracy EDM measurements are planned~\cite{Alarcon:2022ero}, our scenario can be tested in the near future.

\section{Discussions}

In our analysis, we have extended the SM by introducing the dimension-six operator in Eq.\,\eqref{eq:Scp}. 
Within the SM, higher-dimensional operators can be obtained by integrating out heavy fermions such as the top quarks~\cite{Smit:2004kh,Hernandez:2007ng,Garcia-Recio:2009jso,Hernandez:2008db,Brauner:2012gu, Garcia-Recio:2014ffa}. 
These operators are CP-odd due to the CP-violating phases in the quark (and lepton) sector.
One might therefore ask whether sphalerogenesis could be realized via such operators.
However, it has been shown that the CP asymmetry coming from such higher-dimensional operators at $T\sim100\,{\rm GeV}$ is too small~\cite{Brauner:2012gu}. 
Therefore, it is expected that our method using the higher-dimensional operators is not appropriate in the SM. 
Thus, the method developed in Ref.~\cite{Kharzeev:2019rsy} may be appropriate to discuss sphalerogenesis within the framework of the SM. 

\begin{figure}
    \centering
    \includegraphics[width=0.98\linewidth]{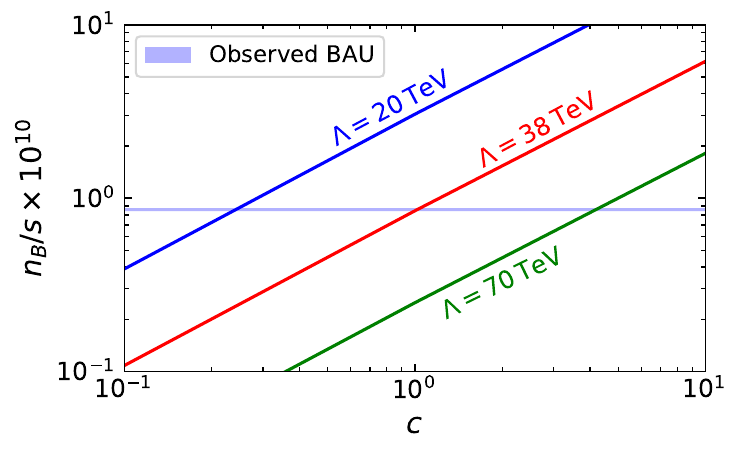}
    \caption{
    The parameter $c$ dependence of the produced baryon asymmetry.
    The blue, red, and green solid lines are the prediction with $\Lambda = 20\,{\rm TeV} \,, 38\,{\rm TeV}$, and $70\,{\rm TeV}$, respectively. 
    }
    \label{fig:nbs_c}
\end{figure}

We briefly comment on the main sources of theoretical uncertainty.
Our reduced model description for the EW sphaleron process assumes the sphaleron ansatz with time-independent profile functions and parametrizes sphaleron-like configurations by a single size parameter $a$. 
More general time dependence and deformations (e.g., $\Xi=a\,\Xi_0,\ \Omega=b\,\Omega_0$ with $a \neq b$) could modify the approximation in Eq.~\eqref{eq:Gamma_sph} and the value of $A_{\rm CP}(a,T)$.
In particular, such contributions modify the decomposition of the lattice rate into sphaleron-like transition rate shown in Eq.~\eqref{eq:Gamma_sph}. 
This source of uncertainty can be partially described by variations in the parameter $c$ in Eq.~\eqref{eq:Gsph_H}, which determines the decoupling temperature for the sphaleron-like processes.
In Figure~\ref{fig:nbs_c}, the $c$-dependence on the produced baryon asymmetry is shown for each $\Lambda$. 
This implies that the prediction for the cutoff $\Lambda$ may contain an $O(1)$ uncertainty if the approximation for Eq.~\eqref{eq:Gamma_sph} has the order-one ambiguity. 
For the uncertainty in $A_{\rm CP}(a,T)$, a quantitative uncertainty budget requires dedicated non-equilibrium lattice simulations of the CP-odd sphaleron process, which is beyond the scope of this work. 
Thus, our results should be viewed as an order-of-magnitude estimate of the baryon asymmetry and its implications for the cutoff scale $\Lambda$.

In general, the dimension-six operator in Eq.\,\eqref{eq:Scp} leads to other interesting predictions such as anomalous triple gauge boson couplings~\cite{Hagiwara:1986vm}. 
The current measurement of the electroweak $Z+2j$ production process constrains $\Lambda > 1.41\,{\rm TeV}$ with the 95\% confidence level~\cite{ATLAS:2020nzk, DasBakshi:2020ejz}. 
In the High-Luminosity LHC (HL-LHC) with the integrated luminosity $3\,{\rm ab}^{-1}$, it is expected that the measurement of the same process gives a constraint $\Lambda > 4.66\,{\rm TeV}$ with the 95\% confidence level~\cite{Hall:2022bme}. 
Using the complementarity between the HL-LHC with lepton colliders such as the International Linear Collider, we may obtain a constraint $\Lambda > 7.64\,{\rm TeV}$~\cite{deBlas:2022ofj}.
These constraints are weaker than those of the electron EDM measurements. 
Thus, it is challenging to test our scenario via anomalous gauge coupling measurements at colliders.

\section{Conclusions}

We have demonstrated the feasibility of sphalerogenesis driven by the dimension-six operator in Eq.\,\eqref{eq:Scp}, which was not considered in Refs.~\cite{Nauta:2000xi, Nauta:2002ru}.
We have found that new physics models that generate this operator with $\Lambda \simeq 38\,{\rm TeV}$ by integrating out heavy fields are able to account for the observed BAU.
Moreover, we have shown that our scenario can be tested by higher-precision electron EDM measurements in the near future.

\section*{Acknowledgment}

MT thanks Takumi Kuwahara for fruitful discussions. 

\bibliographystyle{apsrev4-1}
\bibliography{ref_sphaleron}

\onecolumngrid

\newpage

\appendix

\section{The expression of the sphaleron action \label{app:sphaleron_action}}

We here give the explicit formula of the sphaleron action with the ansatz in Eq.~\eqref{eq:sph_ansatz}. 
Substituting the sphaleron ansatz in Eq.~\eqref{eq:sph_ansatz} into Eq.~\eqref{eq:Ssm} and Eq.~\eqref{eq:Scp}, we obtain 
\begin{align}
\nonumber
S &= \int d^4 x \left[ 
- \frac{1}{2} {\rm tr} \left[ W_{\mu \nu} W^{\mu \nu} \right] + (D_{\mu} \Phi)^{\dagger} D^{\mu} \Phi - V(\Phi) - \frac{g}{3 \Lambda^2} \epsilon_{ijk} \widetilde{W}^{i}_{\mu \nu} W^{j\nu \rho} W^{k\mu}_{\rho} \right] \\
&\equiv \int d\eta \left[ \frac{M(\mu)}{2} \left( \frac{d\mu}{d\eta} \right)^2 + G(\mu) \left( \frac{d\mu}{d \eta} \right)^3 - V(\mu) \right] \,, 
\end{align}
where 
\begin{align}
&\begin{aligned}
M(\mu)
=& ~\frac{2 \pi}{g^2} \int_{0}^{\infty} d \xi \xi^2 
\left[ 
4 \left\{ \frac{4+2c_{\mu}^2}{3} \left( \frac{df}{d\xi} \right)^2 + \frac{4}{\xi^2} \frac{8+2c_{\mu}^2}{3} s_{\mu}^2 f^2 (1-f)^2 \right\} \right. \\ & \left.
+ (1-h)^2 + 2fh(1-f)(1-h) + 2 c_{\mu}^2 f(1-h)^2 \right. \\ & \left. 
+\frac{4+2c_{\mu}^2}{3} \left\{ 
h^2 (1-f)^2 + c_{\mu}^2 f(f-2)(1-h) - 2c_{\mu}^2 fh(1-f)(1-h) 
\right\}
\right] \,, 
\end{aligned}
\\
&\begin{aligned}
G(\mu) 
= \frac{256 \pi}{45} s_{\mu}^2 (4 - s_{\mu}^2) \left(\frac{v}{\Lambda}\right)^2 \,, 
\end{aligned} \\
&\begin{aligned}
V(\mu) 
&= \frac{4 \pi}{g^2} \int_{0}^{\infty} d\xi \xi^2
\left[ 
\frac{4}{\xi^2} s_{\mu}^2  \left\{ \left( \frac{df}{d\xi}\right)^2 + \frac{2}{\xi^2} f^2 (1-f)^2 s_{\mu}^2 \right\} + \frac{s_{\mu}^2}{2} \left( \frac{dh}{d\xi} \right)^2 \right. \\ & \left. 
\quad + \frac{s_{\mu}^2}{\xi^2} \left\{ h^2 (1-f)^2 - 2 c_{\mu}^2 fh (1-f)(1-h) + c_{\mu}^2 f^2(1-h)^2 \right\} + \frac{\lambda}{4g^2}(1-h^2)^2 s_{\mu}^4 \right] \,,
\end{aligned}
\end{align}
with $\xi = gv(T)r$.

\section{Derivation and numerical solutions of the Boltzmann equation \label{sec:BoltzmannEq}}

In this appendix, we describe the explicit expression of the Boltzmann equation given in Eq.~\eqref{eq:boltzmann_eq}. 
As the EW symmetry breaking proceeds, large and small sphaleron-like configurations decouple gradually as shown in Fig.~\ref{fig:Gsph_T}. 
Therefore, the source term $P(T)$ in Eq.\,\eqref{eq:boltzmann_eq} is dominantly produced via such decaying sphaleron-like configurations. 
Thus, $P(T)$ is given by~\cite{Hong:2023zrf}
\begin{align}
\label{eq:PT}
P(T) = \begin{cases} \displaystyle
\int_{a_{\rm min}}^{a_{l}(T)} da \widetilde{\Gamma}_{\rm sph}(a, T) + \int_{a_{u}(T)}^{a_{\rm max}} da \widetilde{\Gamma}_{\rm sph}(a, T) ~~ &(T_{\rm sph} < T < T_{\rm EW}) \\ 
\displaystyle
\Gamma_{\rm lattice}(T) \cdot A_{\rm CP}(a=1, T) ~~ &(T < T_{\rm sph}) 
\end{cases}\,,
\end{align} 
where $\widetilde{\Gamma}_{\rm sph}(a, T) \equiv 3\Gamma_{\rm sph}(a, T) A_{\rm CP}(a, T)$.  
For the value of $a_{\rm min}$ and $a_{\rm max}$, we take $a_{\rm min} =0.001$ and $a_{\rm max} = 5$ as used in Ref.~\cite{Hong:2023zrf}. 
It should be noted that, unlike the result in Ref.\,\cite{Hong:2023zrf}, the factor $A_{\rm CP}(a,T)$ cannot be outside the integral in our formalism because it also depends on the parameter $a$. 

The source term arises during the decay of sphaleron-like configurations that are too large or too small, while the resulting baryon number is washed out by active sphaleron transitions.
For the washout effect term in the Boltzmann equation in Eq.\,\eqref{eq:boltzmann_eq}, it is expressed by~\cite{Hong:2023zrf}
\begin{align}
\label{eq:GammaB}
\Gamma_{B}(T) = \begin{cases} \displaystyle
\frac{39}{4T^3} \int_{a_{\ell}}^{a_{u}} da \Gamma_{\rm sph}(a,T)  &(T_{\rm sph} < T < T_{\rm EW}) \\
\displaystyle
0   &(T < T_{\rm sph})
\end{cases}  \,. 
\end{align}

Solving the Boltzmann equation in Eq.\,\eqref{eq:boltzmann_eq} with Eqs.\,\eqref{eq:Acp_sph}, \eqref{eq:PT} and \eqref{eq:GammaB}, we can estimate a prediction of the baryon asymmetry. 

\begin{figure}[t]
\centering
\includegraphics[width=0.45\linewidth]{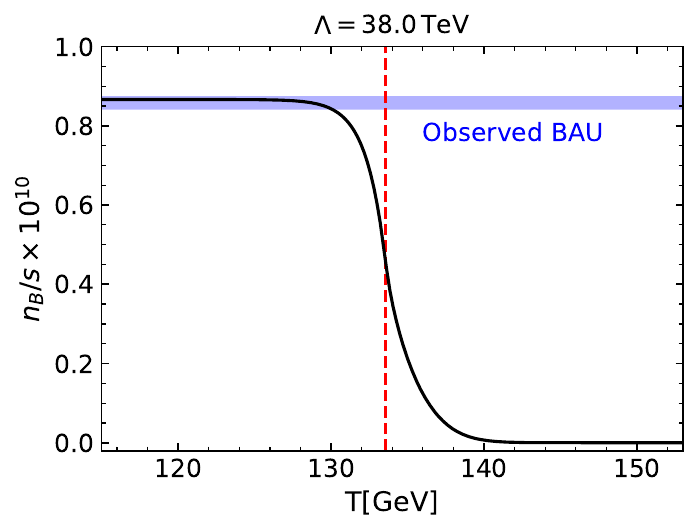}
\caption{
Temperature dependence of $n_{B}/s$ in the case with $\Lambda = 38\,{\rm TeV}$ (black solid line).
The blue region represents the observed value of $n_{B}/s$ shown in Eq.\,\eqref{eq:BAU}. 
The red dashed line is the sphaleron decoupling temperature. 
}
\label{fig:nbs}
\end{figure}

In Fig.\,\ref{fig:nbs}, the numerical result for the prediction of $n_{B}/s$ is shown in the case with $\Lambda =38\,{\rm TeV}$. 
This figure indicates that the baryon asymmetry is produced at $T \simeq T_{\rm sph}$ because the sphaleron washout effect decouples at that temperature. 
In addition, at $T \simeq T_{\rm sph}$, the CP asymmetry of sphaleron-like configuration decay can be large because $A_{\rm CP}(a, T) \propto v(T)^{3/2}$ as confirmed by looking at Eq.\,\eqref{eq:Acp_sph}. 
As a result, the baryon asymmetry is predominantly generated near the sphaleron decoupling temperature $T_{\rm sph}$. 
Moreover, we have numerically confirmed that $A_{\rm CP}(a = 1, T_{\rm sph}) \sim 10^{-6}$, which was predicted to explain the observed BAU in Ref.~\cite{Hong:2023zrf}, is obtained in the parameters used in Fig.\,\ref{fig:nbs}. 

\end{document}